\documentclass[twocolumn]{aastex6}
\usepackage{fixltx2e}
\usepackage{xcolor}
\usepackage{subfigure}
\usepackage{float}
\usepackage{scrextend}
\usepackage{hyperref}
\def\SPSB#1#2{\rlap{\textsuperscript{\textcolor{black}{#1}}}\SB{#2}}

\def\SB#1{\textsubscript{\textcolor{black}{#1}}}
\usepackage{natbib}
\usepackage{wrapfig}
\newcommand\Kepler{\textit{Kepler}}
\shortauthors{Zimmerman et al.}
\shorttitle{Pseudosynchronization of Heartbeat Stars}
 
\begin{document}
\title{The Pseudosynchronization of Binary Stars Undergoing Strong Tidal Interactions}
\author{Mara K. Zimmerman\altaffilmark{1}, Susan E. Thompson\altaffilmark{2, 3}, Fergal Mullally \altaffilmark{2}, Jim Fuller\altaffilmark{4, 5}, Avi Shporer\altaffilmark{4}, Kelly Hambleton\altaffilmark{6} }
\altaffiltext{1}{University of Wyoming 1000 E. University Ave, Laramie, WY 82071}
\altaffiltext{2}{SETI Institute 189 N Bernardo Ave, Mountain View, CA 94043}
\altaffiltext{3}{NASA AMES Reasearch Center Moffett Blvd, Mountain View, CA 94035}
\altaffiltext{4}{California Institute of Technology, 1200 E California Blvd, Pasadena, CA 91125}
\altaffiltext{5}{Kavli Institute for Theoretical Physics, Kohn Hall, University of California, Santa Barbara, CA 93106, USA}
\altaffiltext{6}{Department of Astronomy and Astrophysics, Villanova University, 800 East Lancaster Avenue, Villanova, PA 19085}

\begin{abstract}
Eccentric binaries known as heartbeat stars experience strong dynamical tides as the stars pass through periastron, providing a laboratory to study tidal interactions. We measure the rotation periods of 24 heartbeat systems, using the \Kepler\ light curves to identify rotation peaks in the Fourier transform. Where possible, we compare the rotation period to the pseudosynchronization period derived by \citet{1981Hut}. Few of our heartbeat stars are pseudosynchronized with the orbital period.  For four systems, we were able to identify two sets of rotation peaks, which we interpret as the rotation from both stars in the binary. The majority of the systems have a rotation period that is approximately 3/2 times the pseudosynchronization period predicted by \citet{1981Hut}, suggesting that other physical mechanisms influence the stars' rotation, or that stars typically reach tidal spin equilibrium at a rotation period slightly longer than predicted.

\end{abstract}

\section{Introduction} \label{sec:intro}   
The \Kepler\ Telescope, launched in 2009, has provided four years of near-continuous photometric data of over 150,000 stars in order to look for subtle changes in the brightness of stars \citep{Koch2010}. In addition to finding exoplanets \citep[see e.g.][]{2016Coughlin} and binary stars \citep{2016Kirk}, this mission's data has yielded a type of eccentric binary system known as Heartbeat Binary Stars [HBs]  \citep{2012Thompson, Welsh2011}. These binary stars undergo extreme dynamic tides, causing them to become tidally distorted and more oblate in shape as they pass through periastron. The dynamic tidal distortions, along with heating and Doppler boosting, cause their distinctive pulse-like variations \citep{Welsh2011,2015Hambleton,2016Hambleton}.  Also, these stars show pulsations at frequencies which are harmonics of the orbital frequency, indicating these pulses are likely tidally driven \citep{2002Willems}.

More than 150 heartbeat stars have been identified and cataloged in the Villanova Eclipsing Binary Catalog\footnote{ See also http://keplerebs.villanova.edu/} \citep{2016Kirk}. Accurate and unique models of the photometric variations, as that done for KOI-54 \citep{Welsh2011} and KIC~3749404 \citep{2016Hambleton}, require radial velocity measurements to constrain the possible orbital solutions. From these RV measurements, combined with photometric data, a full orbital solution can be obtained, including the stellar masses, eccentricity and inclination of the system. Recently, radial velocity measurements have been obtained for more than 20 of these binary systems \citep{2015Smullen,2016Shporer} making it possible to model a statistically interesting sample of these tidally active systems.

 Because of the large tidal forces in these HBs, the rotation rates of these stars are expected to quickly pseudosynchronize with the orbital motion \citep{1981Hut}, i.e. the rotation period is comparable to the orbital motion at periastron such that there is no net torque over an orbital cycle. Pseudosynchronization should occur on a timescale approximately ten times faster than the circularization timescale, which is also rapid compared to the lifetime of the star. Many HBs may have reached this state and yet still have eccentric orbits. The precise value of the theoretically expected pseudosynchronous rotation frequency is dependent on both eccentricity and the mechanism of tidal dissipation, and therefore we can test tidal dissipation theories by measuring both orbital eccentricity and the stellar rotation rate. We investigate here if the rotations of the stars have pseudosynchronized with the orbits and whether they are consistent with standard tidal theories.

In this paper, we measure the rotation period of 24~HBs by examining the Fourier transform after removing the orbital heartbeat signal. We then use eccentricity measurements obtained from radial velocity orbital solutions recently reported by \citet{2016Shporer} to determine the pseudosynchronization period of the system. We use the pixel-level \Kepler\ data to rule-out background sources as a likely source of the observed variations.  Finally we compare the rotation and pseudosynchronization rates and discuss our results.

\section{Observations}

We selected HBs from the Villanova Eclipsing Binary Catalog \citep{2016Kirk} and searched for signs of rotation. We particularly focused on those HBs reported in \citet{2012Thompson} and those with radial velocity measurements from \citet{2016Shporer} and \citet{2015Smullen}.

We used the \Kepler\ photometric data from data release 24 \citep{2014Coughlin}. We created our light curves from all 17 \Kepler\ quarters which spanning 1470.5 days, beginning on MJD 54\,953.03 and ending on MJD 54\,423.5\,. We started with the Presearch Data Conditioning (PDC) light curves, which removed common systematic features from the time series light fluctuations using a Bayesian framework \citep{2012Stumpe,2014Stumpe}. We further detrended the data by dividing by low-order polynomials and removing single point outliers.  With the mean-zeroed, relative flux time series we used a Fourier transform method to identify the rotation period of the star. We then looked for evidence of rotation in the residual light curve. 

\begin{figure*}[t]
\centering
\subfigure{\includegraphics[width=.43\textwidth]{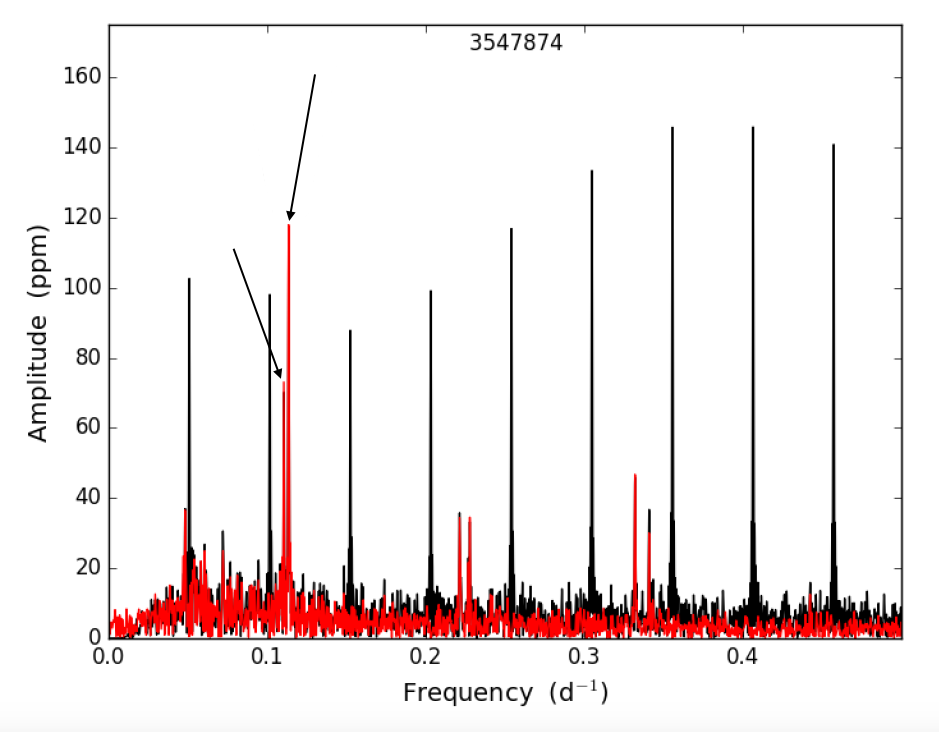}}
\subfigure{\includegraphics[width=.48\textwidth]{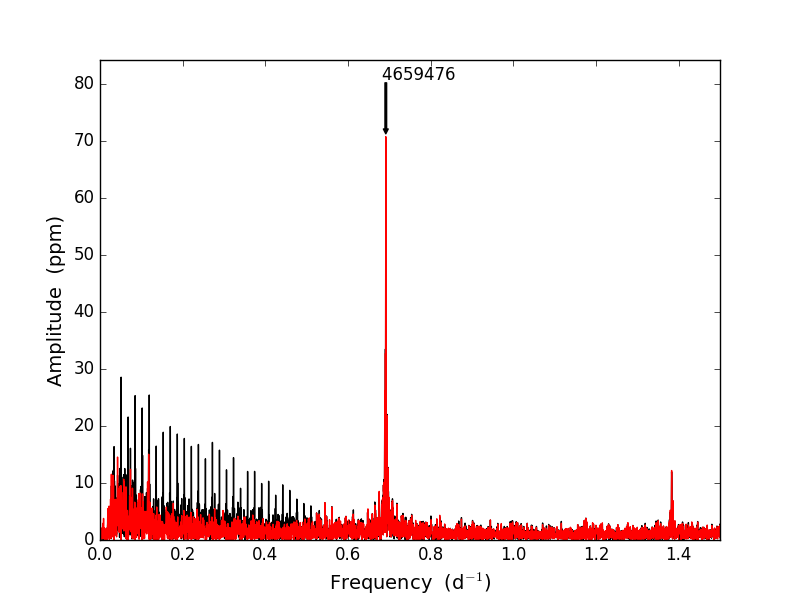}}
\subfigure{\includegraphics[width=.48\textwidth]{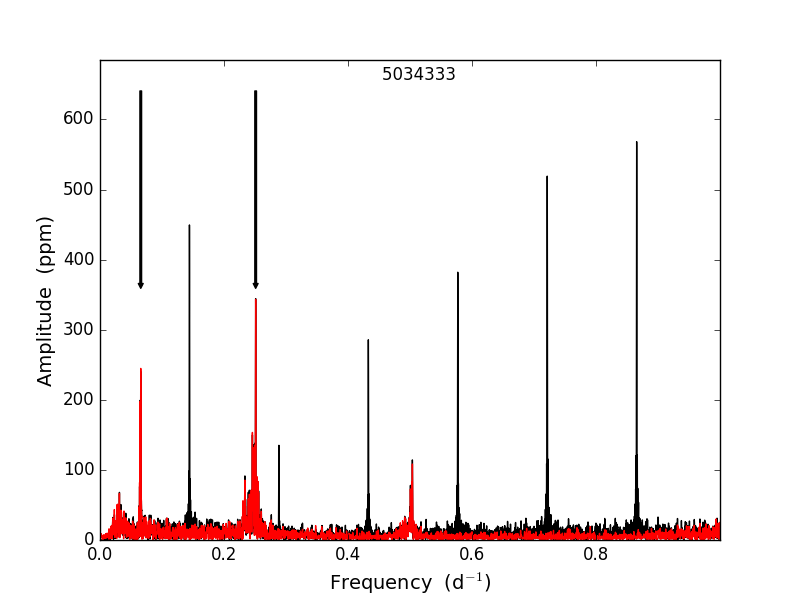}}
\subfigure{\includegraphics[width=.48\textwidth]{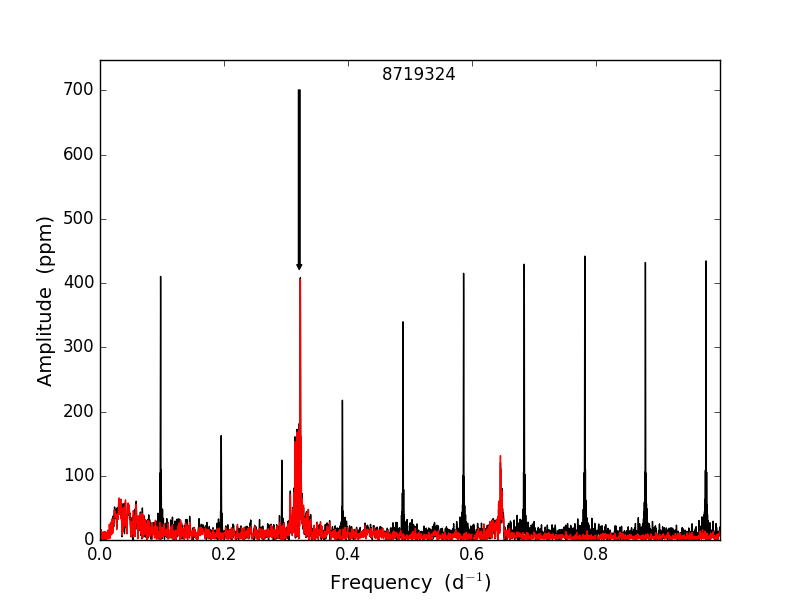}}
\caption{\label{f:ft}  Example Fourier transform of four HBs; the original transform, with no frequencies removed, is in black. The orbital HB frequencies are narrow peaks, equally spaced in frequency. The  Fourier transform after removing these orbital harmonics is overlaid in red. The peaks measured as the rotation frequency is indicated with a black arrow. KIC\,3547874 and KIC\,5034333 have two arrows because they show evidence for two sets of rotation peaks.}
\end{figure*}

\subsection{Rotation Measurements}

\citet{2013McQuillan,2013McQuillanMN} shows that the autocorrelation function (ACF) and a Periodogram give similar measurements of the rotation in most cases. Though the ACF is a more robust and reliable method, in the case of HB systems periodograms can allow to better identify the two sets of periodicities in the light curve, due to the orbit and the stellar rotation. For HB light curves, the ACF methods are confused by the dominant, periodic variations from the orbital motion. However, borrowing from the Periodogram methodology, we find that the Fourier transform can separate the orbital frequencies from those of potential rotation signals.  Because the orbital signal is extremely regular, and not sinusoidal, it appears as a series of evenly spaced harmonic peaks in the Fourier transform.  However, the rotation signals are not strictly periodic and are more sinusoidal. As a result they appear as a group of peaks in the Fourier transform near the same frequency. This is the same type of signal shown in the Periodograms of \Kepler\ observations by \citet{2013McQuillan}. Using this method to measure the rotation rate of \Kepler\ objects, we performed Fourier transforms on the photometric data of stars in our target list.

For each HB, we first removed the periodic pulse caused by the highly eccentric orbit by fitting and removing a sequence of equally spaced frequencies.  We did this by using the tool Period04 \citep{2005Lenz}. We fit the light curve with a series of sine waves at exact harmonics of the orbital period, allowing the amplitude of phase of the harmonics to vary. We then remove these sine-waves from the light curve and visually inspect the residual Fourier transform for evidence of rotation, which would appear as groups of peaks, with at least one following  harmonic group of peaks. We fit the largest peak of this group and report that as the rotation period in Table \ref{tab:rot}. 

In Figure \ref{f:ft}, we show several examples of HB Fourier transforms with the original frequency spectrum in black.  The residuals after removing the harmonics is shown in red to highlight the peaks likely caused by spot modulation. The end result is the measured rotational frequencies. The orbital peaks usually show many distinct, discrete peaks while the rotation peaks are broader than their orbital counterparts but still have at least one harmonic. For four HBs, we found evidence of rotation for both stars in the binary. See, for example, KIC~3547874 and KIC~5034333 in Figure~\ref{f:ft}. In these cases we report both frequencies. 
 
\begin{figure*}[t]
\figurenum{2}
\centering
\subfigure{\includegraphics[width=.49\textwidth]{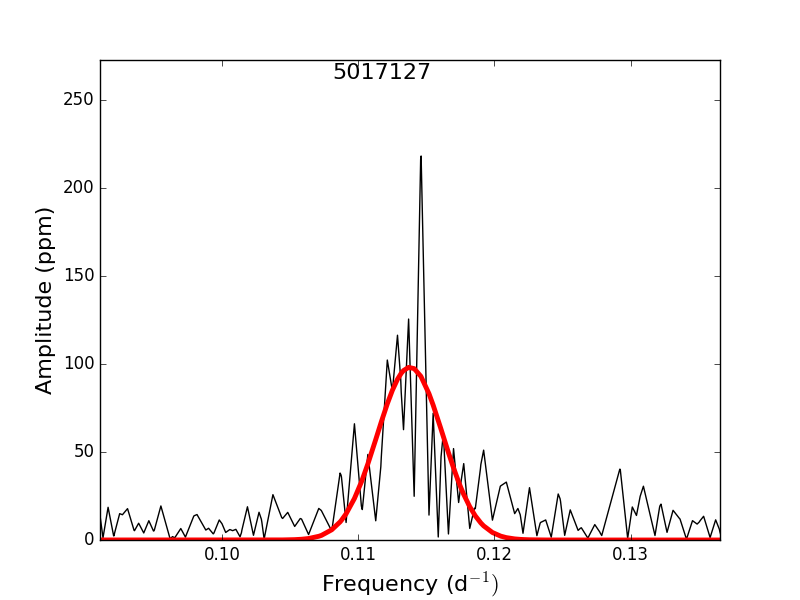}}
\subfigure{\includegraphics[width=.49\textwidth]{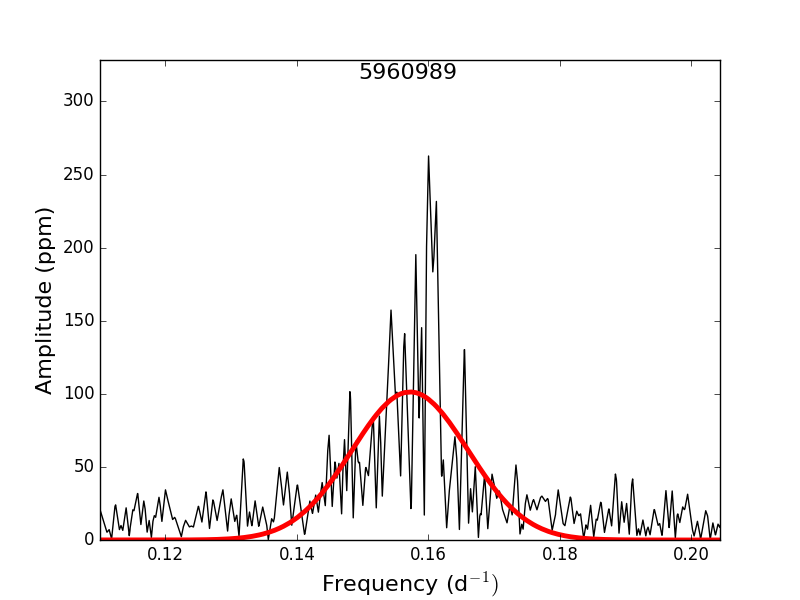}}
\subfigure{\includegraphics[width=.49\textwidth]{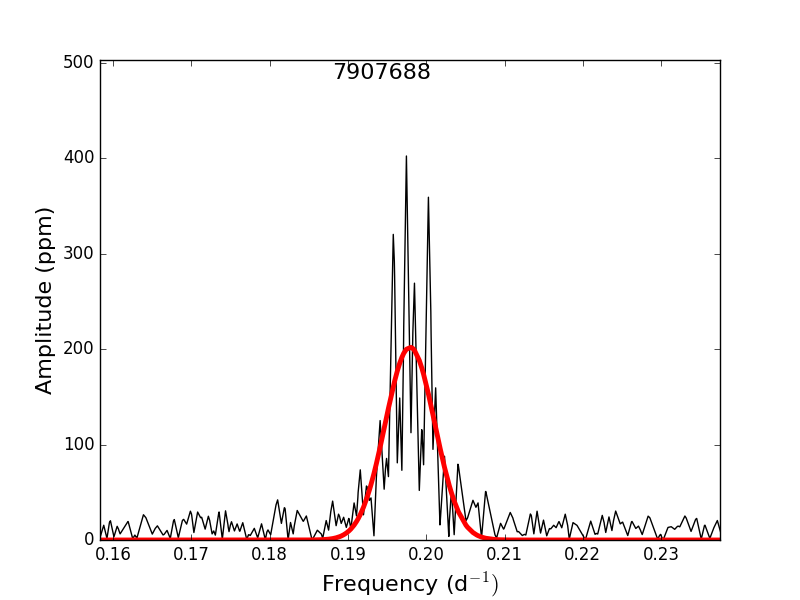}}
\subfigure{\includegraphics[width=.49\textwidth]{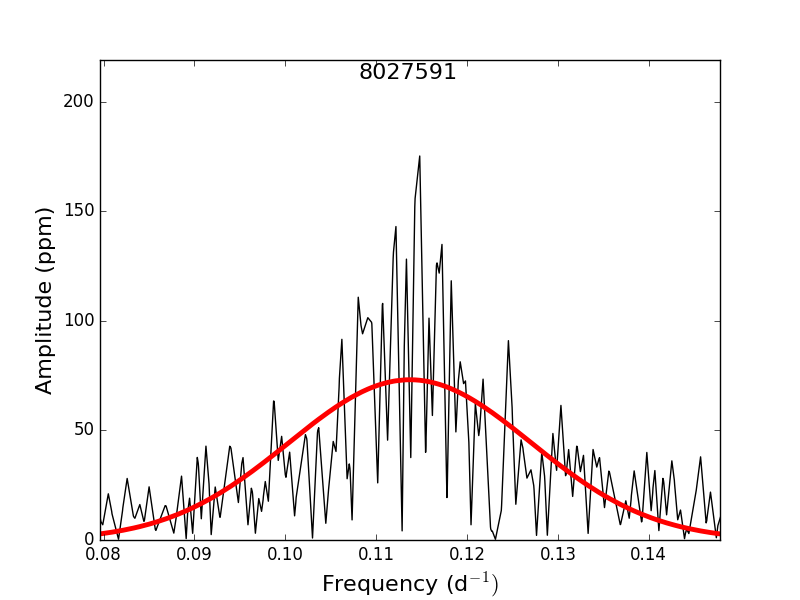}}
\caption{These figures demonstrate the Gaussian fitting used on the Fourier analyses. The black is the original Fourier data, and the red line is the best fit for each of the targets.  \label{fig:Gaussians}}
\end{figure*}

To provide a realistic measure of the error of the rotation period, we fit a Gaussian model to the envelope of peaks that surround the largest rotation peak in the Fourier transform using the package lmfit in Python \citep{2014Newville}. The script optimizes the width, height, and center of the Gaussian peak with least-squares fitting. We fixed the center of the Gaussian curve to the measured rotational frequency, while allowing the height and width of the peak to vary. The code returned the best fit parameters for the Gaussian width which we took as the error in the frequency of the rotation measurement. Examples of several Gaussian fits to the cluster of peaks in the Fourier transform can be seen in Figure \ref{fig:Gaussians}.

\subsection{Ruling-out Obvious Background Sources}
\label{sec:centroids}
We use the measured centroids of our HBs to check for evidence that the star spot signal we see is coming from a nearby star whose pixel response function (PRF) leaks into the target aperture. Periodic changes in the photometric centroid shifts can be caused by periodic changes in the brightness of a nearby star, changes in the spacecraft pointing, or changes in the PRF due to the variations in the telescope focus. Fortunately these latter two occur for the most part on timescales substantially different than the observed stellar period, and can be removed or ignored. The impact of star spots on either the target or a background star can be treated in a similar manner to how \citet{2010Bryson} examined centroid shifts to search for false positive transit signals caused by background eclipsing binary stars. We emphasize two important points from that work: in a crowded field, a centroid shift can be caused by a change in flux in either the target or a background star. However, the lack of a centroid shift is strong evidence that the change in flux is happening on the target star. Therefore, if we see no periodicity in the centroid time-series matching the stellar rotation period, we can conclude the star spot signal is indeed on the target star.

Changes in centroid position due to focus changes over the course of a quarter dwarf all other signals in the centroid position, and we apply four steps to remove them before our analysis. We cut out 96 cadences (approximately 2 days) immediately after any cadence marked as earth-point, and 24 cadences before and after any cadence marked as safe mode, not-fine-point, exclude, or earth-point. The focus typically changes rapidly after an earth-point, leaving a strong signal in the centroid data. Other gaps in the data often show significant trends in the centroids immediately before and after that are also difficult to filter out. See Table 2-3 of the Kepler Archive Manual \citep{2013Thompson}for more details on the meanings of the various data quality flags.  We fit and remove the best fit quadratic to the remaining data in each quarter. We apply a high pass median detrending filter to the residuals, after filling gaps in the data with a cubic polynomial to mitigate edge effects in the filtering process. We sigma clip the remaining data to remove 5$\sigma$ outliers.

\begin{figure}[t]
\figurenum{3}
\includegraphics[width= \linewidth]{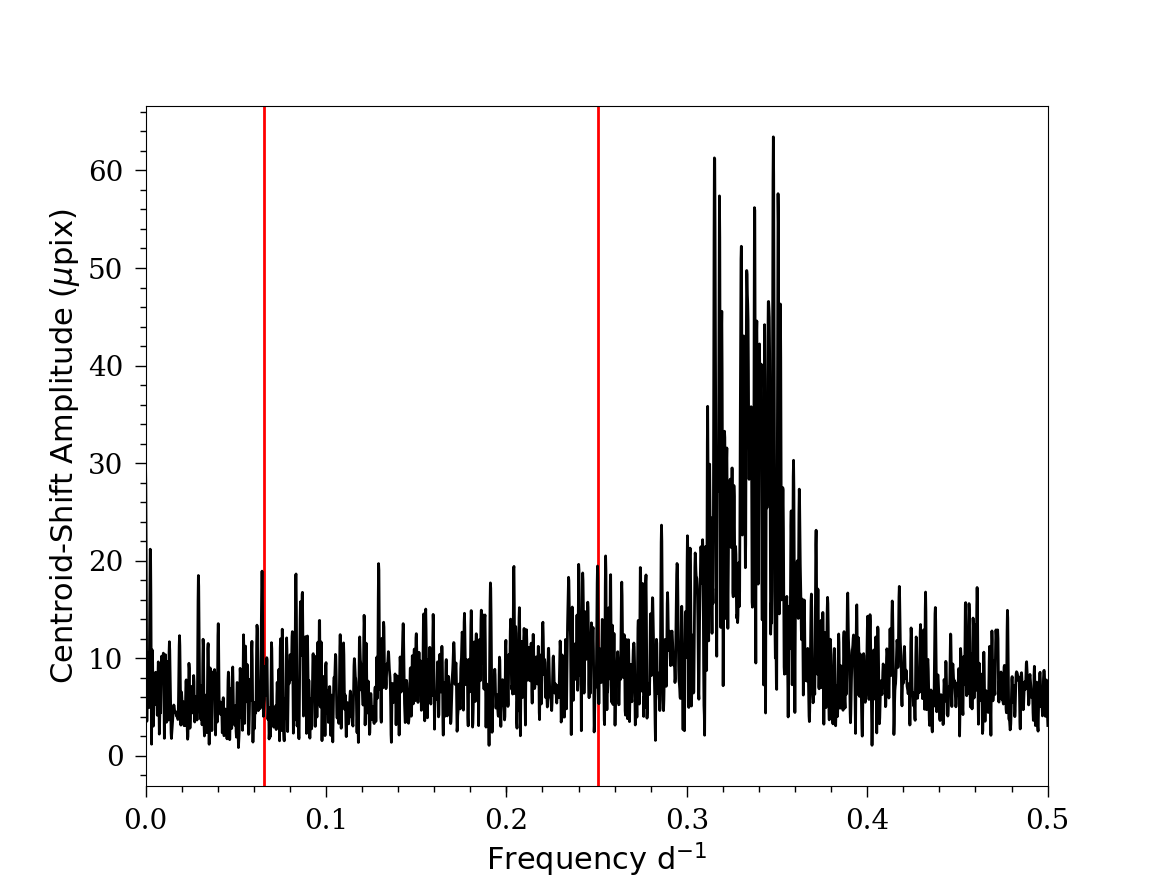}
\caption[width= .5\linewidth]{\label{eg-centroids} A Fourier transform of the centroid position time series for KIC\,5034333. The only significant peaks occur near the 3\,d period of the reaction wheel heater cycle. The red lines indicate the two observed periods due to spots for this star.}
\end{figure}

We apply this process to both the row and column centroid values to quarters 4, 5, 6, and 7 (four consecutive, and relatively well behaved, quarters). We take the FT of both the row and column time series, then plot the quadrature sum of the Fourier transforms. We show and example in Figure~\ref{eg-centroids}. The strongest signal in the resulting spectrum is due to the 3 day focus shift caused by the reaction wheel heater discussed in \S5.2 of the Kepler Data Characteristics Handbook \citet{DCH}. Three of our stars, KIC\,8719324, KIC\,8164262 and KIC\,8775034, have a measured star spot period in this range. We cannot disentangle the reaction wheel heater cycling from possible background star contamination  so we cannot eliminate this possible background star contamination. We conclude that there is no evidence in the centroid data that the star spot signal from any of our other stars is due to contamination from a background star.

\section{The Pseudosynchronization Rate} 
\label{sec:pseudo}
We calculated the pseudosynchronous rotation frequency based on tidal synchronization theory \citet{1981Hut}, which assumes tidal dissipation of the equilibrium tide via a constant parameterized time lag $\tau$. In this theory the pseudosynchronous rotation rate is independent of $\tau$, and is a function of only eccentricity and orbital frequency. The pseudosynchronization rate is determined from a weighted average orbital velocity involving the eccentricity and orbital period. The following equation from \citet{1981Hut} gives the pseudosynchronized rotation rate $\Omega_{ps}$ in terms of the eccentricity $e$ and the orbital frequency $\omega$:

\begin{equation}
\label{eq}
\Omega_{\mathrm{ps}}= \frac{1 +\frac{15}{2}e^2+ \frac{45}{8}e^4+ \frac{5}{16}e^6}{(1-e)^\frac{3}{2}\left(1+3e^3+\frac{3}{8}e^4\right)} \cdot \omega 
\end{equation}

The errors in pseudosynchronization rate were obtained by propagating the error in the eccentricity. In all cases, because we have four years of \Kepler\ data, the orbital frequency is known to a high enough precision to be a negligible portion of the error. Because this equation depends so heavily on eccentricity, we only calculate the pseudosynchronization period for those stars with measured eccentricities. Those from radial velocity measurements are used when available because they are more reliable \citep{2015Smullen, 2016Shporer}. We expand our sample by including the less reliable eccentricities reported from purely photometric fits \citep{2013Thompson} and use a 0.05 error on the eccentricities. The source of the eccentricity measurement is noted in the rightmost column of Table~\ref{tab:rot}.  

\section{Results}

\label{sec:results}
\begin{figure*}[t]
\figurenum{4}
\centering
  \includegraphics[width=0.8\linewidth]{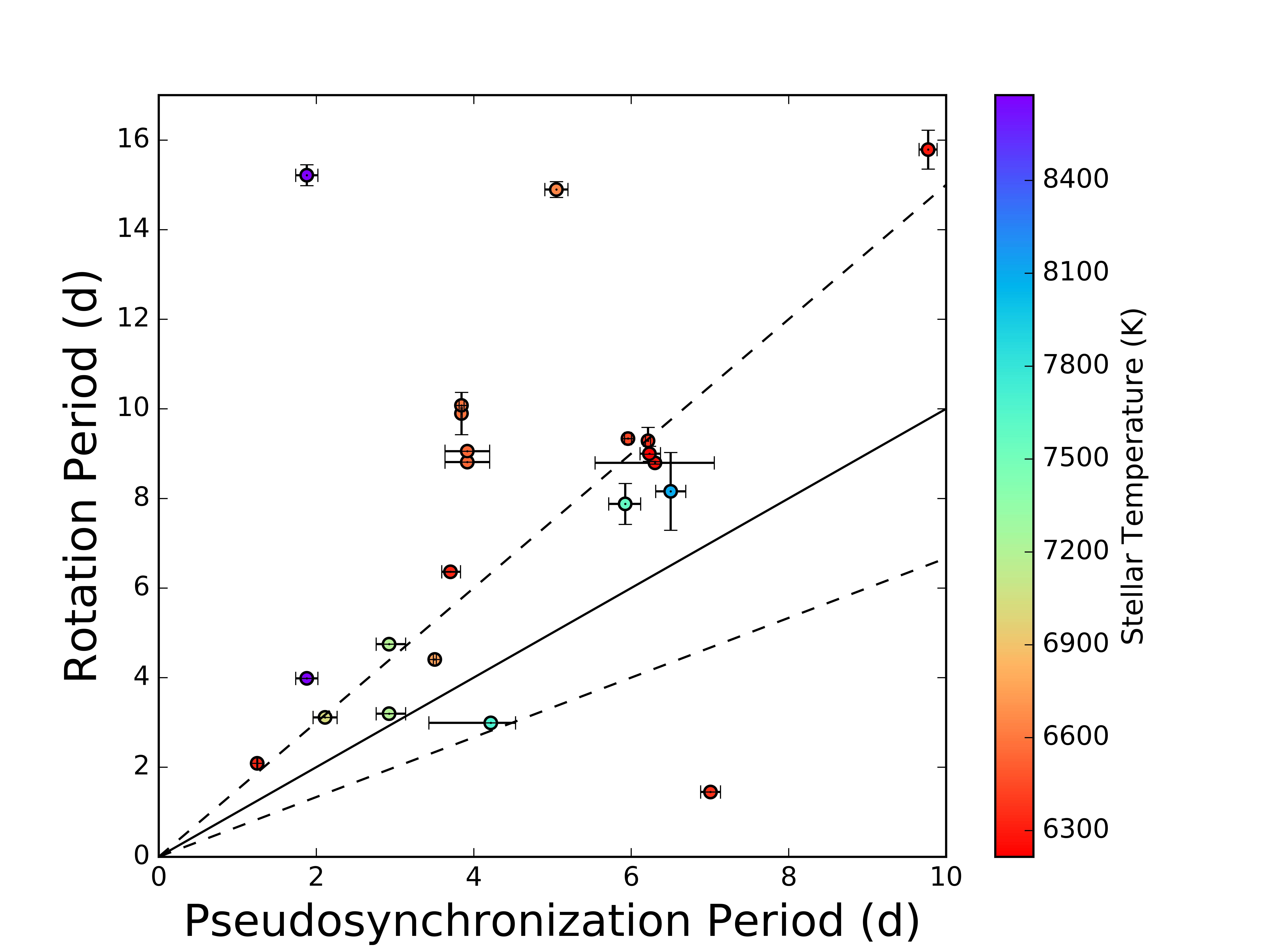}
  \figcaption{The measured rotation period plotted against the calculated pseudosynchonization period. The solid black line shows where the two periods match, i.e. the 1:1 line, while the dashed black lines show the 3:2 and 2:3 lines.
  \\}
  \label{fig:plot}
\end{figure*}

Rotation periods were successfully measured for 24 highly eccentric systems with known heartbeat signals. For 18 of these targets we had eccentricity measurements and were able to compute the pseudosynchronization rate for comparison.  See Table~\ref{tab:rot} for a tabulation of our results and Figure~\ref{fig:plot} for a plot of rotation rate vs pseudosynchronization rate. 

Given the expectation that the rotation should tidally pseudosynchronize, we can split the population into three groups.  1) The majority of the systems have a rotation period that is $\approx 3/2$ times the pseudosychronization period.  2) Four systems have rotation periods significantly longer than that expected from \citet{1981Hut}. Two HBs show a rotation period that is shorter than the pseudosynchronization period. For 4 systems, KIC\,3547874, KIC\,5034333, KIC\,5790807, and KIC\,677504, we found two sets of rotation peaks. In all but one case, KIC\, 5034333, the rotation rates were approximately the same period.  

When looking in the Fourier space for several of these systems, we note that not all peaks can be explained with only the heartbeat signal and the rotation of the stars, lending some doubt on our final interpretation of the Fourier peaks. Systems with unusual features are noted below.

\paragraph{KIC 4659476} The rotation peak for this star has a very large amplitude, however the period is very small (1.446\,d). It is possible that it has two symmetric spots, which would cause us to measure the rotation period at half the true period. However, even then the period is significantly shorter than the pseudosynchronization period. A long period excess exists in the residuals of the Fourier spectrum, but it does not have the typical harmonic signature of rotation.

\paragraph{KIC 5034333} This star shows a strong typical rotation peak at 3.98\,d. See Figure \ref{f:ft}. There is also a significant set of peaks at 15.2\,d, that has only a hint of a harmonic, that we interpret as a rotation peak due to its broad signature and second harmonic. If these are the rotation periods of the two stars in the HB, then one has almost reached pseudosynchronization while the other has not.

\paragraph{KIC 5790807} This star shows two sets of rotation peaks. The frequencies surrounding these peaks were somewhat discrete for a rotation signature, but we fit both peaks and their surrounding noise separately and got two different rotation signatures. These peaks also seem to grow in separation with the second harmonic and then revert to their primary separation at the third harmonic. 

\paragraph{KIC 5818706} This star had two very close peaks in its Fourier transform. However, there were no harmonics associated with the second rotation peak, so we did not include that extra peak in our final measurements. If that peak is caused by spots on the secondary star, it would have a period equal to 9.17\,d, similar to the rotation period we measured for the primary. 

\paragraph{KIC 6775034}  We found two sets of peaks that are likely due to the rotation of the two stars in the heartbeat star system.  The larger amplitude signal has almost pseudosynchronized with the orbit while the lower amplitude signal is near to the 3:2 ratio.  However, a third set of peaks with a harmonic is also visible in the Fourier transform. If this set of peaks was caused by spot rotation, it would indicate a rotation period of 16.29 d, significantly longer than indicated by the other peaks in this transform. 

\section{Discussion}

We know from the Kepler stellar catalog \citep{2017Mathur} and the spectroscopic follow-up of these stars \citep{2016Shporer} that the primary star is usually classified as an AFG-type star. Since the majority of the light originates from the primary star, our measured rotation periods likely originate from the primary star. For stars in the 6,000 - 6,500 K range the typical rotation period measured by \citet{2014McQuillan} was measured to be in the range of 5 - 15 days.  Thus, our measured rotation periods are not much different than in typical F and G type stars. However, none of these HBs, regardless of spectral type, have clearly reached pseudosynchronization. 

Highly eccentric binary systems are expected to synchronize and circularize at a rapid rate because their tidal interactions are efficient at dissipating energy and angular momentum.  The theory of this process is well described in \citet{1977Zahn}, \citet{1981Hut} and later by  \citet{2010Khaliullin} and \citet{2007Khaliullin}. These papers agree that the pseudosynchronization of the rotation should occur well before the orbital circularization.  Given that these systems have not circularized, it has been suggested that some other force, such as a third body, is maintaining this eccentric state in spite of the tidal forces. Evidence that this may be true for HBs can be seen in the rapid apsidal motion seen by \citet{2016Hambleton}. The fact that the rotation has also not synchronized could place strong limits on how long the binary system has lived in this strongly tidal configuration.  

Our rotation period measurements show that the majority of the rotation periods do follow a trend.  Most stars have a rotation period that is $\approx \frac{3}{2}$ times the pseudosynchronization period. Our rotation period measurements show that the majority of the rotation periods are slightly longer than predicted by the pseudosynchronous rotation rate \citet{1981Hut}. One possible reason is that the damping of the equilibrium tide according to the constant lag time model of \citet{1981Hut} is not the dominant tidal torque in these systems. The rotation rate may instead be controlled by the action of dynamical tidal effects such as tidally excited gravity modes, which are indeed observed in some of these systems, e.g., KIC\, 3547874 and KIC\, 50343333. Another possibility is that additional non-tidal torques are important, such as magnetic braking that slows down Sun-like stars with convective envelopes. We expect some level of magnetic braking in our targets since their rotational modulation is presumably the result of magnetic spots created by an active magnetic dynamo \citep{1971Havnes,2017provencal}. If the braking can compete with tidal pseudosynchronization, it could cause the stars to rotate at somewhat longer periods.

Though we interpret the peaks as primary star rotation, there are possible other  physical effects that could be causing us to be measuring another property of the stellar system. Our rotational measurements could be of the secondary and not the primary star. Since the secondary should also be pseudosynchronous, this measurement is still useful and applicable to tidal theory. We could also be measuring the rotation of a third star, either bound or unbound in the HB system, on the same pixel. This would mean the third star contributes significant luminosity to the system, and its radial velocity. The third star would likely have magnetically braked to longer periods, so this would give a measurement much longer than the expected pseudosynchronization \citep{1971Havnes}. Another binary in the same \Kepler\ pixel could cause us to measure ellipsoidal variations in another binary system other than the HB system. The non-HB binary would have to be much fainter than the HB. This is a reasonable possibility for the rotation measurement KIC \, 4659476, given its unusually short rotation period. In this case, the second binary may be quite faint compared to the primary. Since ellipsoidal modulation tends to be larger for shorter period binaries, this would likely cause our measurements to be shorter than expected. 

For targets with RV measurements\citep{2016Shporer,2015Smullen}, we know that the spectral lines were not broadened by very rapid rotation. This could indicate that the primary star does not rotate rapidly, so the peaks in its Fourier transform do not correspond to the primary's rotation. Narrow Fourier peaks with larger amplitude compared to the orbital peakswould point towards indicate a measurement of a source other than rotation. Instead, we could be measuring rotation of a different star, or a close binary on the same Kepler pixel. 

Additionally, the surface of the stars may be differentially rotating. In this case, long-lived spots at different latitudes could produce rotational signals at multiple periods, which would create two or more peaks in the Fourier transform. The discrete peaks in KIC \, 3547874 could be indicative of this differential rotation. In binaries, it is theorized that the tidal friction plays a key role in the efficient synchronization of the system \citep{1984Giuricin}; however, the role must be further examined, given that these HB stars clearly have large dynamic tidal forces but still have not pseudosynchronized. The tidal interactions in these systems can also be affected by the nature and evolution of the star: in stars with convective envelopes, turbulent tidal friction can affect the equilibrium tide, while in stars with radiative convective envelopes, radiative damping can also be responsible for slowing the tidal dynamics \citep{1977Zahn}.

Those systems that lie much above the synchronization line are a bit of a mystery. They suggest that either the strong tidal forces are insufficient to change to rotation periods, or that they have been in this eccentric configuration for too little time to allow for the tides to synchronize the rotation. The tidally induced pulsations in these stars can also play a role in affecting the orbital and rotational dynamics \citep{2013Hambleton}. 

The odd star KIC\,4659476 has an unexpected unusually short rotation rate. While in most cases, this could also be caused by contamination or a background source, it is unusual for stars to rotate at such a short period, so contamination from another source is less likely, though still a reasonable possibility. The stellar parameters for this star indicate that it is a subgiant: the \Kepler\ stellar catalog \citep{2017Mathur}, using the granulation driven light curve method called 'flicker' \citep{2016Bastien} to obtain an accurate log(g), gives a stellar temperature of 6384\,K, log(g) of 3.972 and a mass of 1.31 solar masses. One type of rapidly rotating subgiant is known as an FK~Comae variable and is believed to be caused by merging a W~UMa contact binary system \citep{1981Bopp}. If this system was previously a triple star system and is now an FK~Comae variable in an eccentric binary, we may expect this target to be X-ray bright or show chromospheric activity \citep{2016Howell}.

\section{Conclusions}
We successfully measured rotation for 24 HBs; for 18 HBs we also found the predicted pseudosynchronization rate using tidal synchronization theory \citep{1981Hut}. As is seen in Figure \ref{fig:plot}, the rotation of these noteworthy stars do not follow the expected pattern. Instead, these stars are clustered around roughly $\frac{3}{2}$ times the predicted rotation rate, indicating that they have plateaued prematurely in their synchronization. This could also indicate that the pseudosynchronous rotation rate predicted by the constant time lag model of \citet{1981Hut}  does not hold true for HB systems, perhaps because other effects, such as dynamical tides or magnetic braking, more greatly influence the star's evolution.

As more HBs with photometrically determined rotation periods are found with survey missions like K2 \citep{2014Howell}, TESS [Transiting Exoplanet Survey Satellite] \citep{TESS}, and PLATO [PLAnetary Transits and Oscillations of stars] \citep{Plato} we will be able to further explore the strong dynamic tidal forces that are influencing the evolution of these binary systems.

\floattable
\begin{deluxetable}{ccccccc}[p]
\tablenum{1}
\tablecaption{Parameters for Heartbeat Stars \label{tab:rot}}
\tablewidth{0pt}
\tablehead{
\colhead{KIC} & \colhead{Orbital Period} & \colhead{Rotation Period} & \colhead{Eccentricity} & \colhead{Temperature} & \colhead{Pseudosynchronization} & \colhead{References}
\\
\colhead{ } & \colhead{(d)} & \colhead{(d)} & \colhead{}  & \colhead{K} &\colhead{(d)} &\colhead{ }}
\startdata
3547874 A	&	19.6921722	(	765	)	&	8.8100	(	26	)	&	0.648	\SPSB{	0.050	}	{	0.050	}	&	6549	&	3.92	\SPSB{	0.28	}	{	0.28	}	&	1	\\
3547874 B	&	19.6921722	(	765	)	&	9.05	(	91	)	&	0.648	\SPSB{	0.050	}	{	0.050	}	&	6549	&	3.92	\SPSB{	0.28	}	{	0.28	}	&	1	\\
4659476	&	58.9963737	(	368	)	&	1.446	(	14	)	&	0.745	\SPSB{	0.0110	}	{	0.0110	}	&	6376	&	7.01	\SPSB{	0.13	}	{	0.13	}	&	2	\\
4949187	&	11.9773917	(	380	)	&	5.194	(	52	)	&	-							&	-	&	-							&		\\
5017127	&	20.0064041	(	780	)	&	9.34	(	11	)	&	0.550	\SPSB{	0.0050	}	{	0.0050	}	&	6443	&	5.96	\SPSB{	0.05	}	{	0.05	}	&	2	\\
5034333 A	&	6.9322800	(	170	)	&	3.98	(	60	)	&	0.570	\SPSB{	0.050	}	{	0.050	}	&	8675	&	1.88	\SPSB{	0.14	}	{	0.14	}	&	1	\\
5034333 B	&	6.9322800	(	170	)	&	15.2	(	12	)	&	0.570	\SPSB{	0.050	}	{	0.050	}	&	8675	&	1.88	\SPSB{	0.14	}	{	0.14	}	&	1	\\
5090937	&	8.8006929	(	240	)	&	8.16	(	51	)	&	0.241	\SPSB{	0.0130	}	{	0.0130	}	&	8092	&	6.50	\SPSB{	0.19	}	{	0.19	}	&	2	\\
5790807 A	&	79.9962462	(	543	)	&	9.9	(	10	)	&	0.857	\SPSB{	0.0031	}	{	0.0030	}	&	6582	&	3.84	\SPSB{	0.01	}	{	0.01	}	&	2	\\
5790807 B	&	79.9962462	(	543	)	&	10.1	(	11	)	&	0.857	\SPSB{	0.0031	}	{	0.0030	}	&	6582	&	3.84	\SPSB{	0.03	}	{	0.03	}	&	2	\\
5818706	&	14.9599406	(	514	)	&	9.29	(	17	)	&	0.454	\SPSB{	0.0038	}	{	0.0039	}	&	6378	&	6.21	\SPSB{	0.03	}	{	0.03	}	&	2	\\
5960989	&	50.7215338	(	298	)	&	6.4	(	15	)	&	0.813	\SPSB{	0.0170	}	{	0.0150	}	&	6331	&	3.70	\SPSB{	0.11	}	{	0.13	}	&	2	\\
6775034 A	&	10.0285473	(	294	)	&	3.185	(	32	)	&	0.556	\SPSB{	0.0470	}	{	0.0370	}	&	7187	&	2.92	\SPSB{	0.17	}	{	0.21	}	&	2	\\
6775034 B	&	10.0285473	(	294	)	&	4.745	(	31	)	&	0.556	\SPSB{	0.0470	}	{	0.0370	}	&	7187	&	2.92	\SPSB{	0.17	}	{	0.21	}	&	2	\\
7041856	&	4.0006688	(	771	)	&	1.7407	(	17	)	&	-							&	-	&	-							&		\\
7259722	&	9.6332256	(	275	)	&	17.3	(	94	)	&	-							&	-	&	-							&		\\
7907688	&	4.3448368	(	886	)	&	5.05	(	25	)	&	-							&	-	&	-							&		\\
8027591	&	24.2744321	(	103	)	&	8.79	(	43	)	&	0.586	\SPSB{	0.0082	}	{	0.0083	}	&	6279	&	6.30	\SPSB{	0.76	}	{	0.75	}	&	2	\\
8164262	&	87.4571700	(	638	)	&	2.99	(	38	)	&	0.857	\SPSB{	0.0260	}	{	0.0650	}	&	7700	&	4.22	\SPSB{	0.79	}	{	0.31	}	&	2	\\
8719324	&	10.2326979	(	300	)	&	3.11	(	19	)	&	0.640	\SPSB{	0.050	}	{	0.050	}	&	7023	&	2.11	\SPSB{	0.15	}	{	0.15	}	&	3	\\
9965691	&	15.6831951	(	553	)	&	12.32	(	86	)	&	-							&	-	&	-							&		\\
10162999	&	3.4292146	(	622	)	&	2.086	(	63	)	&	0.473	\SPSB{	0.0032	}	{	0.0032	}	&	6335	&	1.25	\SPSB{	0.01	}	{	0.01	}	&	2	\\
11071278	&	55.8852250	(	329	)	&	9.00	(	13	)	&	0.755	\SPSB{	0.0150	}	{	0.0130	}	&	6215	&	6.23	\SPSB{	0.12	}	{	0.14	}	&	2	\\
11240948	&	3.4019372	(	615	)	&	1.554	(	16	)	&	-							&	-	&	-							&		\\
11403032	&	7.6316344	(	197	)	&	14.89	(	22	)	&	0.288	\SPSB{	0.0130	}	{	0.0130	}	&	6657	&	5.05	\SPSB{	0.15	}	{	0.15	}	&	2	\\
11649962	&	10.5627371	(	312	)	&	4.4	(	73	)	&	0.521	\SPSB{	0.0035	}	{	0.0035	}	&	6762	&	3.51	\SPSB{	0.02	}	{	0.02	}	&	2	\\
11923629	&	17.9732836	(	673	)	&	15.79	(	15	)	&	0.363	\SPSB{	0.0058	}	{	0.0059	}	&	6293	&	9.77	\SPSB{	0.12	}	{	0.11	}	&	2	\\
12255108	&	9.1315263	(	253	)	&	7.88	(	35	)	&	0.296	\SPSB{	0.0150	}	{	0.0160	}	&	7577	&	5.92	\SPSB{	0.21	}	{	0.20	}	&	2	\\
\enddata
\tablecomments{\label{tab} \textbf{References} All orbital periods were obtained from \citet{2016Kirk}. The references column indicates the source of the eccentricity value; the sources are as follows (1) \citet{2012Thompson} (2) \citet{2016Shporer} (3) \citet{2015Smullen} The reported temperatures are from the \Kepler catalog; they are reported in this table and in \ref{fig:plot} for reference.}
\end{deluxetable}

\bibliography{hb}

\begin{thebibliography}{}
\expandafter\ifx\csname natexlab\endcsname\relax\def\natexlab#1{#1}\fi

\bibitem[{{ Newville} {et~al.}(2014){ Newville}, {Stensitzki}, {Allen}, \&
  {Ingargiola}}]{2014Newville}
{ Newville}, M., {Stensitzki}, T., {Allen}, D.~B., \& {Ingargiola}, A. 2014,
  Zendo, 1001.0331

\bibitem[{{Bastien} {et~al.}(2016){Bastien}, {Stassun}, {Basri}, \&
  {Pepper}}]{2016Bastien}
{Bastien}, F.~A., {Stassun}, K.~G., {Basri}, G., \& {Pepper}, J. 2016, \apj,
  818, 43

\bibitem[{{Bopp} \& {Stencel}(1981)}]{1981Bopp}
{Bopp}, B.~W., \& {Stencel}, R.~E. 1981, \apjl, 247, L131

\bibitem[{{Bryson} {et~al.}(2010){Bryson}, {Tenenbaum}, {Jenkins},
  {Chandrasekaran}, {Klaus}, {Caldwell}, {Gilliland}, {Haas}, {Dotson}, {Koch},
  \& {Borucki}}]{2010Bryson}
{Bryson}, S.~T., {Tenenbaum}, P., {Jenkins}, J.~M., {et~al.} 2010, \apjl, 713,
  L97

\bibitem[{{Coughlin} {et~al.}(2014){Coughlin}, {Thompson}, {Bryson}, {Burke},
  {Caldwell}, {Christiansen}, {Haas}, {Howell}, {Jenkins}, {Kolodziejczak},
  {Mullally}, \& {Rowe}}]{2014Coughlin}
{Coughlin}, J.~L., {Thompson}, S.~E., {Bryson}, S.~T., {et~al.} 2014, \aj, 147,
  119

\bibitem[{{Coughlin} {et~al.}(2016){Coughlin}, {Mullally}, {Thompson}, {Rowe},
  {Burke}, {Latham}, {Batalha}, {Ofir}, {Quarles}, {Henze}, {Wolfgang},
  {Caldwell}, {Bryson}, {Shporer}, {Catanzarite}, {Akeson}, {Barclay},
  {Borucki}, {Boyajian}, {Campbell}, {Christiansen}, {Girouard}, {Haas},
  {Howell}, {Huber}, {Jenkins}, {Li}, {Patil-Sabale}, {Quintana}, {Ramirez},
  {Seader}, {Smith}, {Tenenbaum}, {Twicken}, \& {Zamudio}}]{2016Coughlin}
{Coughlin}, J.~L., {Mullally}, F., {Thompson}, S.~E., {et~al.} 2016, \apjs,
  224, 12

\bibitem[{{Giuricin} {et~al.}(1984){Giuricin}, {Mardirossian}, \&
  {Mezzetti}}]{1984Giuricin}
{Giuricin}, G., {Mardirossian}, F., \& {Mezzetti}, M. 1984, \aap, 131, 152

\bibitem[{{Hambleton} {et~al.}(2015){Hambleton}, {Kurtz}, {Pr{\v s}a},
  {Fuller}, \& {Thompson}}]{2015Hambleton}
{Hambleton}, K., {Kurtz}, D., {Pr{\v s}a}, A., {Fuller}, J., \& {Thompson}, S.
  2015, in {European Physical Journal Web of Conferences}, Vol. 101, {European
  Physical Journal Web of Conferences}, 04007

\bibitem[{{Hambleton} {et~al.}(2013){Hambleton}, {Degroote}, {Conroy},
  {Bloemen}, {Kurtz}, {Thompson}, {Fuller}, {Giammarco}, {Pablo}, \& {Pr{\v
  s}a}}]{2013Hambleton}
{Hambleton}, K., {Degroote}, P., {Conroy}, K., {et~al.} 2013, in EAS
  Publications Series, Vol.~64, EAS Publications Series, ed. K.~{Pavlovski},
  A.~{Tkachenko}, \& G.~{Torres}, 285--294

\bibitem[{{Hambleton} {et~al.}(2016){Hambleton}, {Kurtz}, {Pr{\v s}a}, {Quinn},
  {Fuller}, {Murphy}, {Thompson}, {Latham}, \& {Shporer}}]{2016Hambleton}
{Hambleton}, K., {Kurtz}, D.~W., {Pr{\v s}a}, A., {et~al.} 2016, \mnras, 463,
  1199

\bibitem[{{Havnes} \& {Conti}(1971)}]{1971Havnes}
{Havnes}, O., \& {Conti}, P.~S. 1971, \aap, 14, 1

\bibitem[{{Howell} {et~al.}(2016){Howell}, {Mason}, {Boyd}, {Smith}, \&
  {Gelino}}]{2016Howell}
{Howell}, S.~B., {Mason}, E., {Boyd}, P., {Smith}, K.~L., \& {Gelino}, D.~M.
  2016, \apj, 831, 27

\bibitem[{{Howell} {et~al.}(2014){Howell}, {Sobeck}, {Haas}, {Still},
  {Barclay}, {Mullally}, {Troeltzsch}, {Aigrain}, {Bryson}, {Caldwell},
  {Chaplin}, {Cochran}, {Huber}, {Marcy}, {Miglio}, {Najita}, {Smith},
  {Twicken}, \& {Fortney}}]{2014Howell}
{Howell}, S.~B., {Sobeck}, C., {Haas}, M., {et~al.} 2014, \pasp, 126, 398

\bibitem[{{Hut}(1981)}]{1981Hut}
{Hut}, P. 1981, \aap, 99, 126

\bibitem[{{Khaliullin} \& {Khaliullina}(2007)}]{2007Khaliullin}
{Khaliullin}, K.~F., \& {Khaliullina}, A.~I. 2007, \mnras, 382, 356

\bibitem[{{Khaliullin} \& {Khaliullina}(2010)}]{2010Khaliullin}
---. 2010, \mnras, 401, 257

\bibitem[{{Kirk} {et~al.}(2016){Kirk}, {Conroy}, {Pr{\v s}a}, {Abdul-Masih},
  {Kochoska}, {Matijevi{\v c}}, {Hambleton}, {Barclay}, {Bloemen}, {Boyajian},
  {Doyle}, {Fulton}, {Hoekstra}, {Jek}, {Kane}, {Kostov}, {Latham}, {Mazeh},
  {Orosz}, {Pepper}, {Quarles}, {Ragozzine}, {Shporer}, {Southworth},
  {Stassun}, {Thompson}, {Welsh}, {Agol}, {Derekas}, {Devor}, {Fischer},
  {Green}, {Gropp}, {Jacobs}, {Johnston}, {LaCourse}, {Saetre}, {Schwengeler},
  {Toczyski}, {Werner}, {Garrett}, {Gore}, {Martinez}, {Spitzer}, {Stevick},
  {Thomadis}, {Vrijmoet}, {Yenawine}, {Batalha}, \& {Borucki}}]{2016Kirk}
{Kirk}, B., {Conroy}, K., {Pr{\v s}a}, A., {et~al.} 2016, \aj, 151, 68

\bibitem[{{Koch} {et~al.}(2010){Koch}, {Borucki}, {Basri}, {Batalha}, {Brown},
  {Caldwell}, {Christensen-Dalsgaard}, {Cochran}, {DeVore}, {Dunham},
  {Gautier}, {Geary}, {Gilliland}, {Gould}, {Jenkins}, {Kondo}, {Latham},
  {Lissauer}, {Marcy}, {Monet}, {Sasselov}, {Boss}, {Brownlee}, {Caldwell},
  {Dupree}, {Howell}, {Kjeldsen}, {Meibom}, {Morrison}, {Owen}, {Reitsema},
  {Tarter}, {Bryson}, {Dotson}, {Gazis}, {Haas}, {Kolodziejczak}, {Rowe}, {Van
  Cleve}, {Allen}, {Chandrasekaran}, {Clarke}, {Li}, {Quintana}, {Tenenbaum},
  {Twicken}, \& {Wu}}]{Koch2010}
{Koch}, D.~G., {Borucki}, W.~J., {Basri}, G., {et~al.} 2010, \apjl, 713, L79

\bibitem[{{Lenz} \& {Breger}(2005)}]{2005Lenz}
{Lenz}, P., \& {Breger}, M. 2005, Communications in Asteroseismology, 146, 53

\bibitem[{Mathur {et~al.}(2017)Mathur, Huber, Batalha, Ciardi, Bastien,
  Bieryla, Buchhave, Cochran, Endl, Esquerdo, Furlan, Howard, Howell, Isaacson,
  Latham, MacQueen, \& Silva}]{2017Mathur}
Mathur, S., Huber, D., Batalha, N.~M., {et~al.} 2017, The Astrophysical Journal
  Supplement Series, 229, 30

\bibitem[{{McQuillan} {et~al.}(2013{\natexlab{a}}){McQuillan}, {Aigrain}, \&
  {Mazeh}}]{2013McQuillanMN}
{McQuillan}, A., {Aigrain}, S., \& {Mazeh}, T. 2013{\natexlab{a}}, \mnras, 432,
  1203

\bibitem[{{McQuillan} {et~al.}(2013{\natexlab{b}}){McQuillan}, {Mazeh}, \&
  {Aigrain}}]{2013McQuillan}
{McQuillan}, A., {Mazeh}, T., \& {Aigrain}, S. 2013{\natexlab{b}}, \apjl, 775,
  L11

\bibitem[{{McQuillan} {et~al.}(2014){McQuillan}, {Mazeh}, \&
  {Aigrain}}]{2014McQuillan}
---. 2014, \apjs, 211, 24

\bibitem[{{Provencal} {et~al.}(2017){Provencal}, {Hermes}, {Kawaler},
  {Shipman}, {Bischoff-Kim}, \& {Thompson}}]{2017provencal}
{Provencal}, J.~L., {Hermes}, J.~J., {Kawaler}, S.~K., {et~al.} 2017, in
  Astronomical Society of the Pacific Conference Series, Vol. 509, 20th
  European White Dwarf Workshop, ed. P.-E. {Tremblay}, B.~{Gaensicke}, \&
  T.~{Marsh}, 359

\bibitem[{{Rauer} {et~al.}(2014){Rauer}, {Catala}, {Aerts}, {Appourchaux},
  {Benz}, {Brandeker}, {Christensen-Dalsgaard}, {Deleuil}, {Gizon}, {Goupil},
  {G{\"u}del}, {Janot-Pacheco}, {Mas-Hesse}, {Pagano}, {Piotto}, {Pollacco},
  {Santos}, {Smith}, {Su{\'a}rez}, {Szab{\'o}}, {Udry}, {Adibekyan}, {Alibert},
  {Almenara}, {Amaro-Seoane}, {Eiff}, {Asplund}, {Antonello}, {Barnes},
  {Baudin}, {Belkacem}, {Bergemann}, {Bihain}, {Birch}, {Bonfils}, {Boisse},
  {Bonomo}, {Borsa}, {Brand{\~a}o}, {Brocato}, {Brun}, {Burleigh}, {Burston},
  {Cabrera}, {Cassisi}, {Chaplin}, {Charpinet}, {Chiappini}, {Church},
  {Csizmadia}, {Cunha}, {Damasso}, {Davies}, {Deeg}, {D{\'{\i}}az}, {Dreizler},
  {Dreyer}, {Eggenberger}, {Ehrenreich}, {Eigm{\"u}ller}, {Erikson}, {Farmer},
  {Feltzing}, {de Oliveira Fialho}, {Figueira}, {Forveille}, {Fridlund},
  {Garc{\'{\i}}a}, {Giommi}, {Giuffrida}, {Godolt}, {Gomes da Silva},
  {Granzer}, {Grenfell}, {Grotsch-Noels}, {G{\"u}nther}, {Haswell}, {Hatzes},
  {H{\'e}brard}, {Hekker}, {Helled}, {Heng}, {Jenkins}, {Johansen},
  {Khodachenko}, {Kislyakova}, {Kley}, {Kolb}, {Krivova}, {Kupka}, {Lammer},
  {Lanza}, {Lebreton}, {Magrin}, {Marcos-Arenal}, {Marrese}, {Marques},
  {Martins}, {Mathis}, {Mathur}, {Messina}, {Miglio}, {Montalban}, {Montalto},
  {Monteiro}, {Moradi}, {Moravveji}, {Mordasini}, {Morel}, {Mortier},
  {Nascimbeni}, {Nelson}, {Nielsen}, {Noack}, {Norton}, {Ofir}, {Oshagh},
  {Ouazzani}, {P{\'a}pics}, {Parro}, {Petit}, {Plez}, {Poretti}, {Quirrenbach},
  {Ragazzoni}, {Raimondo}, {Rainer}, {Reese}, {Redmer}, {Reffert},
  {Rojas-Ayala}, {Roxburgh}, {Salmon}, {Santerne}, {Schneider}, {Schou},
  {Schuh}, {Schunker}, {Silva-Valio}, {Silvotti}, {Skillen}, {Snellen}, {Sohl},
  {Sousa}, {Sozzetti}, {Stello}, {Strassmeier}, {{\v S}vanda}, {Szab{\'o}},
  {Tkachenko}, {Valencia}, {Van Grootel}, {Vauclair}, {Ventura}, {Wagner},
  {Walton}, {Weingrill}, {Werner}, {Wheatley}, \& {Zwintz}}]{Plato}
{Rauer}, H., {Catala}, C., {Aerts}, C., {et~al.} 2014, Experimental Astronomy,
  38, 249

\bibitem[{{Ricker} {et~al.}(2014){Ricker}, {Winn}, {Vanderspek}, {Latham},
  {Bakos}, {Bean}, {Berta-Thompson}, {Brown}, {Buchhave}, {Butler}, {Butler},
  {Chaplin}, {Charbonneau}, {Christensen-Dalsgaard}, {Clampin}, {Deming},
  {Doty}, {De Lee}, {Dressing}, {Dunham}, {Endl}, {Fressin}, {Ge}, {Henning},
  {Holman}, {Howard}, {Ida}, {Jenkins}, {Jernigan}, {Johnson}, {Kaltenegger},
  {Kawai}, {Kjeldsen}, {Laughlin}, {Levine}, {Lin}, {Lissauer}, {MacQueen},
  {Marcy}, {McCullough}, {Morton}, {Narita}, {Paegert}, {Palle}, {Pepe},
  {Pepper}, {Quirrenbach}, {Rinehart}, {Sasselov}, {Sato}, {Seager},
  {Sozzetti}, {Stassun}, {Sullivan}, {Szentgyorgyi}, {Torres}, {Udry}, \&
  {Villasenor}}]{TESS}
{Ricker}, G.~R., {Winn}, J.~N., {Vanderspek}, R., {et~al.} 2014, in \procspie,
  Vol. 9143, Space Telescopes and Instrumentation 2014: Optical, Infrared, and
  Millimeter Wave, 914320

\bibitem[{{Shporer} {et~al.}(2016){Shporer}, {Fuller}, {Isaacson}, {Hambleton},
  {Thompson}, {Pr{\v s}a}, {Kurtz}, {Howard}, \& {O'Leary}}]{2016Shporer}
{Shporer}, A., {Fuller}, J., {Isaacson}, H., {et~al.} 2016, \apj, 829, 34

\bibitem[{{Smullen} \& {Kobulnicky}(2015)}]{2015Smullen}
{Smullen}, R.~A., \& {Kobulnicky}, H.~A. 2015, \apj, 808, 166

\bibitem[{{Stumpe} {et~al.}(2014){Stumpe}, {Smith}, {Catanzarite}, {Van Cleve},
  {Jenkins}, {Twicken}, \& {Girouard}}]{2014Stumpe}
{Stumpe}, M.~C., {Smith}, J.~C., {Catanzarite}, J.~H., {et~al.} 2014, \pasp,
  126, 100

\bibitem[{{Stumpe} {et~al.}(2012){Stumpe}, {Smith}, {Van Cleve}, {Twicken},
  {Barclay}, {Fanelli}, {Girouard}, {Jenkins}, {Kolodziejczak}, {McCauliff}, \&
  {Morris}}]{2012Stumpe}
{Stumpe}, M.~C., {Smith}, J.~C., {Van Cleve}, J.~E., {et~al.} 2012, \pasp, 124,
  985

\bibitem[{{Thompson} {et~al.}(2012){Thompson}, {Everett}, {Mullally},
  {Barclay}, {Howell}, {Still}, {Rowe}, {Christiansen}, {Kurtz}, {Hambleton},
  {Twicken}, {Ibrahim}, \& {Clarke}}]{2012Thompson}
{Thompson}, S.~E., {Everett}, M., {Mullally}, F., {et~al.} 2012, \apj, 753, 86

\bibitem[{{Thompson} {et~al.}(2013){Thompson}, {Hambleton}, {Mullally},
  {Christiansen}, {Everett}, {Howell}, {Barclay}, {Burke}, {Still}, \&
  {Rowe}}]{2013Thompson}
{Thompson}, S.~E., {Hambleton}, K., {Mullally}, F., {et~al.} 2013, in American
  Astronomical Society Meeting Abstracts, Vol. 221, American Astronomical
  Society Meeting Abstracts \#221, 142.05

\bibitem[{{Van Cleve} \& {Christiansen}(2016)}]{DCH}
{Van Cleve}, J.~E., \& {Christiansen}, J.~L. 2016, KSCI - 19040-005

\bibitem[{{Welsh} {et~al.}(2011){Welsh}, {Orosz}, {Aerts}, {Brown},
  {Brugamyer}, {Cochran}, {Gilliland}, {Guzik}, {Kurtz}, {Latham}, {Marcy},
  {Quinn}, {Zima}, {Allen}, {Batalha}, {Bryson}, {Buchhave}, {Caldwell},
  {Gautier}, {Howell}, {Kinemuchi}, {Ibrahim}, {Isaacson}, {Jenkins}, {Prsa},
  {Still}, {Street}, {Wohler}, {Koch}, \& {Borucki}}]{Welsh2011}
{Welsh}, W.~F., {Orosz}, J.~A., {Aerts}, C., {et~al.} 2011, \apjs, 197, 4

\bibitem[{{Willems} \& {Aerts}(2002)}]{2002Willems}
{Willems}, B., \& {Aerts}, C. 2002, \aap, 384, 441

\bibitem[{{Zahn}(1977)}]{1977Zahn}
{Zahn}, J.-P. 1977, \aap, 57, 383

\end{thebibliography}
\end{document}